\documentstyle[12pt,epsfig]{article}
\title{ Radiative Leptonic Decays of $B_c $ Meson }

\author{ Chao-Hsi Chang$^{a,b}$, Jian-Ping Cheng$^c$ and
Cai-Dian  L\"{u}$^d$\footnote{Alexander von
Humboldt foundation fellow.}\\
{\small a CCAST (World Laboratory), P.O. Box 8730,
Beijing 100080, China;}\\
{\small b Institute of Theoretical Physics, Academia  Sinica, P.O.Box
2735, Beijing 100080, China;}\\
{\small c Deutsches Elektronen Synchrotron (DESY), Notkestrasse 85,
D-22607 Hamburg, Germany;}\\
{\small d II Institut f\"ur Theoretische Physik, Universit\"at Hamburg,
D-22761 Hamburg, Germany}
}

\date{}
\begin{document}

\maketitle
\begin{picture}(0,0)(0,0)
\put(340,290){{\large DESY 97-246}}
\put(340,270){{\large AS-ITP-97-027}}
\put(340,250){{\large December, 1997}}
\end{picture}

\begin{abstract}
To the leading order,
the radiative leptonic decays $B_c\to\gamma\ell \bar\nu$ (  $ \ell=e,
\mu$ ) are studied carefully.
In the study, a non-relativistic constituent
quark model
and the effective Lagrangian for the heavy flavour decays
are used. As a result, the branching ratios
turn out to be of the orders of $10^{-5}$ for $B_c\to \gamma
\mu \bar \nu $ or  for $B_c \to \gamma e \bar \nu $.
Based on the study, we point out
the decays being accessible experimentally at the future LHC,
and the possibility to determine the decay constant $f_{B_c}$
through the radiative decays.

\end{abstract}
\newpage

\section{Introduction}

The pure-leptonic decays of heavy mesons are
very interesting from the point of not only theoretical
but also experimental view \cite{cch1,1}. In 
principle, the pure-leptonic
decays $\bar{B_c} \to \ell \bar \nu$ can be
used to determine the decay constant $f_{B_c}$.
It is also discussed that the pure-leptonic decays
of $B_c$ meson may be sensitive
to new physics beyond the Standard Model (SM) at tree level \cite{1}.
Based on the estimates in \cite{cch,2},
except at Tevatron, numerous $B_c$ mesons 
are hard to be produced in the current colliders,
whereas at LHC a great number of $B_c$ mesons, e.g.
$2\times 10^8$, may be produced.
Thus we may expect that careful experimental studies
on the $B_c$ meson will be able to be accessible
in the foreseeable future.
In addition, the $B_c$ meson decay channels can also contribute
some background
for probing $B^\pm$ decays with the same final states \cite{bpm},
thus, it makes an `extra' reason to study these decay channels of the $B_c$
meson precisely.

The decays of pseudoscalar mesons
into light lepton pairs are helicity suppressed,
i.e. their decay widths are suppressed by $m_\ell^2/m_B^2$:
\begin{equation}
\Gamma (B_c \to \ell \bar \nu )=\frac{G_F^2}{8\pi^2} |V_{cb}|^2
f_{B_c}^2 m_{B_c}^3 \frac{m_\ell^2}{m_{B_c}^2}
\left( 1-  \frac{m_\ell^2}{m_{B_c}^2} \right) ^2.
\end{equation}
Thus it is hard to collect enough events of the pure-leptonic decays
i.e. it is very difficult to obtain very good statistics for the decays,
so that it makes very difficult to determine the decay constant
$f_{B_c}$ from these processes.
In SM, only the decay $B_c\to\tau \bar \nu_\tau$ does not suffer so much
from this suppression and the branching ratio can be about 1.5\%
\cite{cch1}.
However the produced $\tau$ decays promptly and
at least one more neutrino is generated when the cascade decay
is taken into account, thus this decay channel is 
difficult to be identified.
The experimental efficiency should be discussed in connection
with a specific detector if one insists on using the
$\tau$ leptonic decay for the purpose to determine 
the decay constant $f_{B_c}$.

In the present work, we will
study the processes $B_c\to\gamma\ell\bar\nu$
within SM and with the effective Lagrangian
for the heavy flavour decays.
In the following section
we will analyze $B_c\to\gamma\ell\bar\nu$ in the framework
of a constituent quark model. Finally we will
discuss the obtained results
briefly in the last section.

\section{Model calculations }

Because of the lightness of the leptons $e$ and $\mu$, the processes
$ B_c \to \ell \bar \nu$ are suppressed  much
by the helicity factor $m_l^2/m_{B_c}^2$ as in eqn.(1).
If an additional photon line is attached to any of the charged lines
of the Feynman diagrams for the pure-leptonic decays as done
in Fig.1, i.e. the pure-leptonic
processes change into the corresponding radiative ones.
The situation will be different: now no helicity suppression 
exists any more, but there will be an additional $\alpha$ (the 
electro-magnetic coupling constant)
suppression instead.
To the radiative decays there are four tree diagrams to contribute,
as shown in Fig. 1.
It is easy to see that
the fourth diagram (Fig.1d), in which the
photon is emitted from the W boson,
is suppressed further by a factor of $m_b^2/m_W^2$,
if comparing it with
the other three diagrams. Thus we neglect it for simplicity.
To be consistent, in the following calculations we will neglect
all the terms suppressed by this factor $m_b^2/m_W^2$.
It is easy to check that the total amplitudes is gauge invariant
at this accuracy. Hence to the accuracy, the amplitudes
corresponding to the other three diagrams turn out to be
\begin{eqnarray}
{\cal H}_{a+b} &=& -i \sqrt{2} G_F e V_{cb} \bar c \left 
[  Q_c \not \!
\epsilon_\gamma \frac{ \not \!p
_\gamma -\not \! p_c +m_c}{(p_c \cdot p_\gamma)} \gamma_\mu P_L+ Q_b P_R
\gamma_\mu \frac{\not \! p_b -\not \! p_\gamma  +m_b}{(p_b\cdot p_\gamma)}
\not \! \epsilon_\gamma \right] b ~(\bar \ell \gamma ^\mu P_L \nu) ,
\nonumber\\
{\cal H}_{c} &=& -i \sqrt{2} G_F e  V_{cb} (\bar c \gamma ^\mu P_L b)
~ \left [ \bar \ell \not \!
\epsilon_\gamma \frac{ \not \!p
_\gamma +\not \! p_\ell +m_\ell }{(p_\ell \cdot p_\gamma)} \gamma_\mu P_L
\nu \right] .
\label{h4}
\end{eqnarray}
\begin{figure}
  \begin{center}
    \epsfig{file=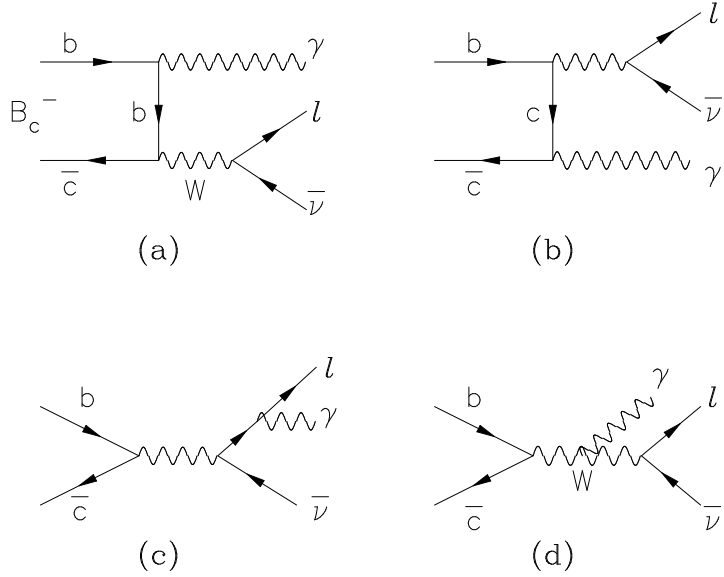,bbllx=1cm,bblly=9.5cm,bburx=20cm,bbury=19.5cm,%
width=14cm,angle=0}
    \caption{Feynman diagrams in standard model for $B_c \to  \gamma \ell
 \bar \nu$}
    \label{f1}
  \end{center}
\end{figure}

To be at the ``quark level'',
the amplitudes given in eqn.(\ref{h4}) are not sufficient enough to analyze
the processes $B_c \to \gamma \ell \bar \nu $,
indeed at least a model is needed to `turn' the amplitudes into
the `hadronic level'.
Here for the aim we adopt a simple  constituent quark model
(see, for example \cite{cheng}).
In this model both of the quark and anti-quark inside the meson
are treated non-relativistically moving with
the same velocity and the quark masses are the constituent masses.
In the constituent quark model, we have
\begin{equation}
p_c^\mu=(m_c /m_{B_c})p_{B_c}^\mu, ~~~~~
p_b^\mu=(m_b /m_{B_c})p_{B_c}^\mu. \label{s2}
\end{equation}

We use further the interpolating field technique \cite{hqet}
which relates the hadronic matrix elements
to the decay constants of the mesons in the present case.
With the decay constant definition:
\begin{equation}
<0|\bar q \gamma^\mu \gamma_5 b|B_c> = i f_{B_c} p_B^\mu,\label{dd}
\end{equation}
the whole amplitude for $B_c \to \gamma \ell \bar \nu $ decay is derived from
eqn.~(\ref{h4},\ref{s2}):
\begin{eqnarray}
{\cal A} &=& \frac{\sqrt{2} e G_FV_{cb}}{6(p_{B_c} \cdot p_\gamma)}
 f_{B_c} 
\left[ \left(\frac{m_{B_c}}{m_b}-2\frac{m_{B_c}}{m_c}\right) i
 \epsilon_{\mu\nu  \alpha \beta } p_{B_c}^\nu p_\gamma^\alpha
\epsilon_\gamma^\beta\right. \nonumber \\
&&\left.+\left(6- \frac{m_{B_c}}{m_b}-2\frac{m_{B_c}}{m_c}\right)
( p_{\gamma\nu}
\epsilon_{\gamma\mu} -p_{\gamma \mu}\epsilon_{\gamma\nu})p_{B_c}^\nu \right
]
(\bar \ell \gamma ^\mu P_L \nu).\label{5}
\end{eqnarray}

Note once again that in the above calculations, all the
terms suppressed by the factor of $m_\ell/m_b$
have been neglected.
It is easily seen that the eqn.(\ref{5}) is explicitly gauge invariant.
 Neglecting the mass of the lepton, we get the
differential decay width:
\begin{equation}
\frac{d\Gamma}{d \hat s d \hat t} 
=\frac{\alpha  G_F^2 |V_{cb}|^2  }{144\
pi^2 }
f_{B_c}^2 m_{B_c}^3 \frac{\hat s}{ (1-\hat 
s)^2}\left[ x_b (1-\hat s-\hat t
)^2
+x_c \hat t^2 \right]  ,\label{6}
\end{equation}
where 
$$x_b=\left( 3-\frac{m_{B_c}}{m_b}\right)^2,~~~~~
x_c=\left(3-2\frac{m_{B_c}}{m_c}\right)^2.$$
The $\hat s$, $\hat t$ are defined as $ \hat s =(p_\ell + p_\nu)^2/m_{B_c
}^2$,
$ \hat t =(p_\ell + p_\gamma)^2/m_{B_c}^2$.
Hence the decay width is:
\begin{equation}
\Gamma =\frac{ \alpha G_F^2 |V_{cb}|^2  }{2592\pi^2 }
f_{B_c}^2 m_{B_c}^3 \left[ x_b+x_c \right].
\label{7}
\end{equation}
Using $\alpha=1/132$,
 $|V_{cb} |=0.04$ \cite{pdg}, $m_c=1.5$ GeV, $m_{B_c}=6.258$ GeV 
\cite{eq},
we obtain
\begin{equation}
\Gamma (B_c \to \gamma \ell \bar \nu )=6.2\times 10^{-17} \times
\left(\frac{f_{B_c} }{360MeV}\right)^2~{\rm GeV}. \label{9}
\end{equation}
If the lifetime is taken as $\tau(B_c)=0.52\times 10^{-12} s$ \cite{life},
and the decay constant is used as $f_{B_c}=360$MeV \cite{fbc},
the branching ratio is found to be $4.9\times 10^{-5}$.

Since it is helpful for experiments to detect this decay channel,
it is also useful to consider the differential spectra.
The photon energy spectrum is easily derived from eqn.(\ref{6}):
\begin{equation}
  \label{eg}
  \frac{m_B}{\Gamma}\frac{d\Gamma}{dE_\gamma} =
\frac{1}{\Gamma}\frac{d\Gamma}{d\lambda_\gamma}=24 \lambda_\gamma
(1-2\lambda_\gamma),
\end{equation}
with $\lambda_\gamma = E_\gamma/m_B$.
We show the photon energy spectrum in Fig.2 as the solid line.
This is clearly different from the bremsstrahlung photon spectrum,
but the same with $B^\pm \to \gamma \ell \nu $ decay photon spectrum 
\cite{bu}.
\begin{figure}[htbp]
  \begin{center}
   \epsfig{file=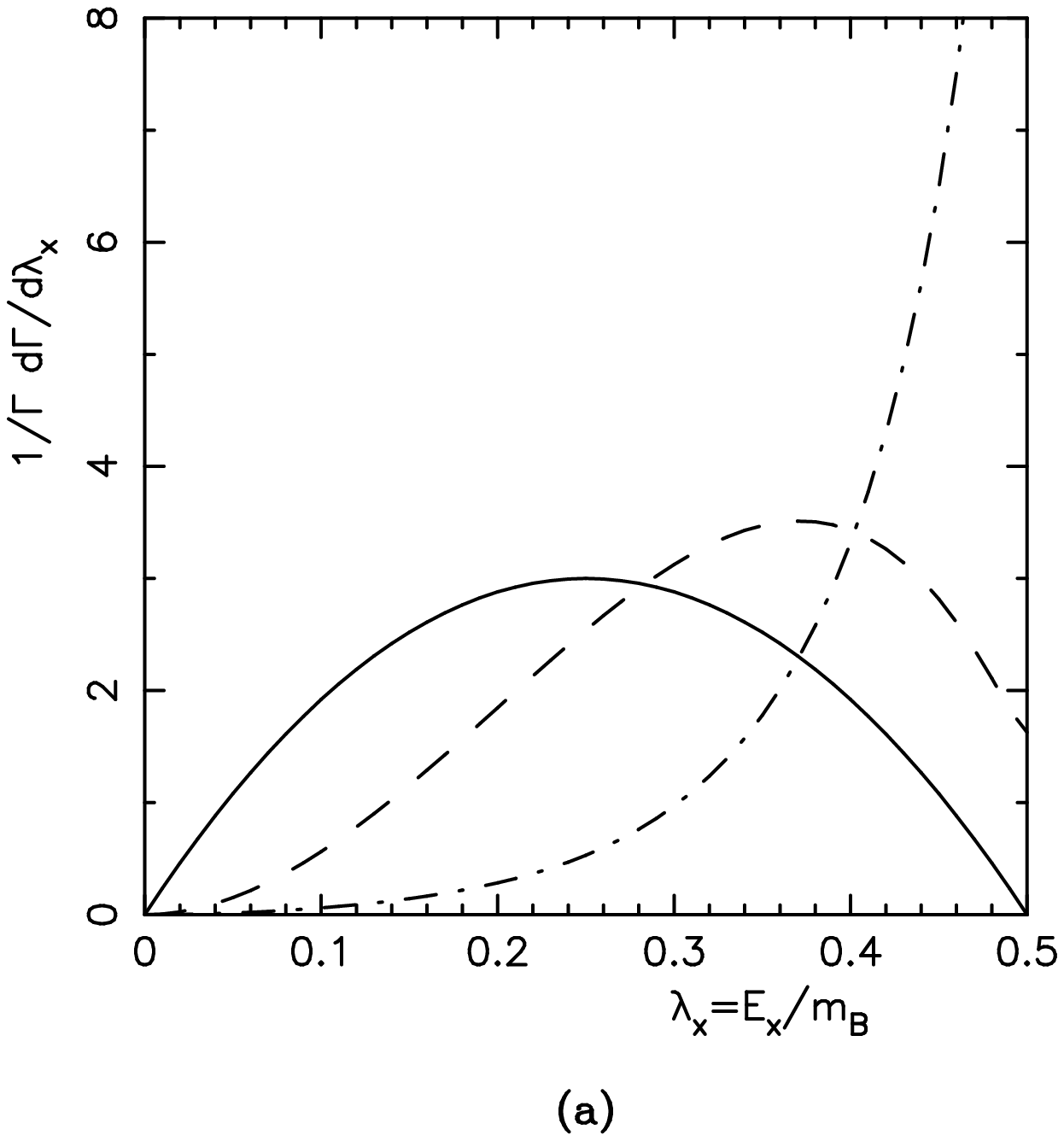,bbllx=5cm,bblly=6.5cm,bburx=18cm,bbury=17cm,%
width=8cm,angle=0}
 \epsfig{file=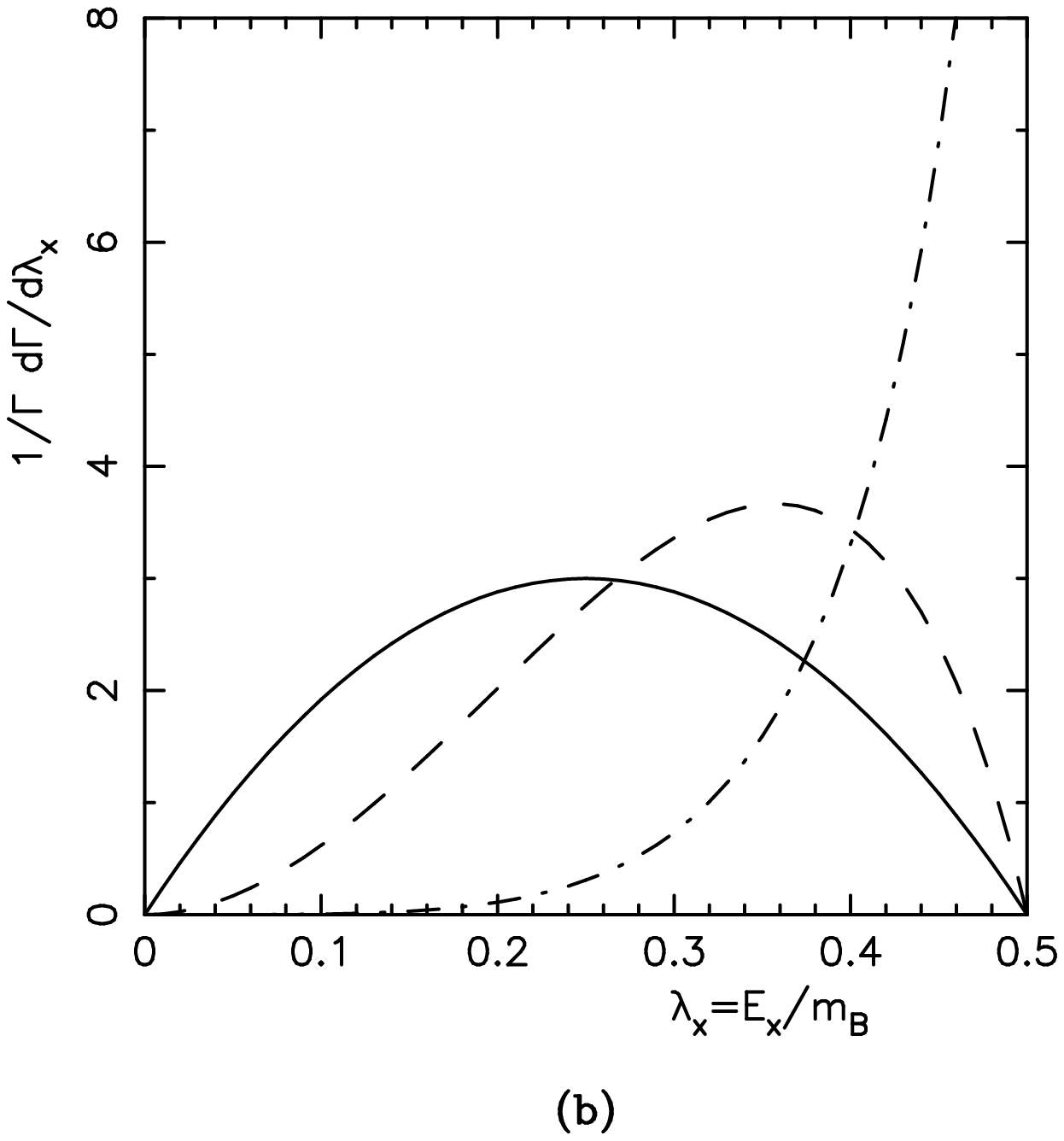,bbllx=5cm,bblly=6.5cm,bburx=18cm,bbury=17cm,%
width=8cm,angle=0}
     \caption{Normalized energy spectra of the decay
$B_c \to \gamma \ell \bar \nu$ (a) and $B_u \to \gamma \ell \bar \nu$ (b). 
The solid line is for the photon energy
spectrum, the dashed line is for the neutrino energy and the dash-dotted 
line
is for the lepton energy spectrum, respectively.}
    \label{f2}
  \end{center}
\end{figure}

The invariant mass $t$ of the charged lepton and photon combination is
directly related to the energy carried by the neutrino, i.e.
the missing energy in the process, so it is also measurable.
Thus to present the charged lepton energy distribution
and  as well as the neutrino one are interesting. They are
\begin{eqnarray}
  \label{en}
\frac{1}{\Gamma}\frac{d\Gamma}{d\lambda_\nu}&=&\frac{36}{x_b+x_c}
\left\{ x_c(1-2\lambda_\nu)\left[ 2\lambda_\nu +(1-2\lambda_\nu)
\ln(1-2\lambda_\nu)\right] \right.\\\nonumber
&&~~~~+ \left.
x_b \left[ 2\lambda_\nu(3-5\lambda_\nu) +(1-2\lambda_\nu) (3-2\lambda_\nu)
\ln(1-2\lambda_\nu)\right]\right\},
\end{eqnarray}
\begin{eqnarray}
  \label{el}
\frac{1}{\Gamma}\frac{d\Gamma}{d\lambda_\ell}&=&\frac{36}{x_b+x_c}
\left\{ x_b(1-2\lambda_\ell)\left[ 2\lambda_\ell +(1-2\lambda_\ell)
\ln(1-2\lambda_\ell)\right]  \right.\\\nonumber
&&~~~~+  \left.
x_c \left[ 2\lambda_\ell(3-5\lambda_\ell) +(1-2\lambda_\ell) (3-2\lambda_\ell)
\ln(1-2\lambda_\ell)\right]\right\},
\end{eqnarray}
where  $\lambda_\nu = E_\nu/m_B$, $\lambda_\ell = E_\ell/m_B$.
Since both c quark  and b quark inside $B_c$ meson
are heavy, the contributions
from the terms proportional to $x_c$, $x_b$, 
both are important.

To show the distributions more clearly,
we present $ \frac{1}{\Gamma}\frac{d\Gamma}{d\lambda_x}, (x=\gamma,\nu,\ell)$
in Fig.2(a) as solid, dashed and dash-dotted lines.
In  $B^\pm \to \gamma \ell \nu $ decay, since the lightness of the u quark,
$x_u >> x_b$, the energy spectra for charged lepton and neutrino are kept
only the $x_c$ terms in (\ref{en},\ref{el}).
They are shown in Fig.2(b) correspondingly.
The behavior is similar, but the endpoint at $\lambda_i=0.5$ is different
2E
for $B_c\to \gamma\ell\bar \nu$,
$$\left .\frac{1}{\Gamma} \frac{d \Gamma}{d\lambda_\nu }
\right |_{\lambda_\nu=0.5 }=1.6,~~~~
\left .\frac{1}{\Gamma} \frac{ d \Gamma}{d\lambda_\ell }
\right |_{\lambda_\ell=0.5 }=16;$$
while for $B^\pm\to \gamma\ell \nu$,
$$\left .\frac{1}{\Gamma} \frac{d \Gamma}{d\lambda_\nu }
\right|_{\lambda_\nu=0.5} =0,~~~~~~
\left .\frac{1}{\Gamma} \frac{ d \Gamma}{d\lambda_\ell }
\right |_{\lambda_\ell=0.5} =18.$$

For comparison, the pure-leptonic decay branching ratios are also given
with the same parameters as the radiative ones:
\begin{eqnarray}
B(B_c\to\mu \bar \nu_\mu)&=&6.2\times 10^{-5},\nonumber\\
 B(B_c\to e \bar \nu_e)&=&1.4\times 10^{-9}.
\end{eqnarray}
It is easy to see that the branching ratios of
the radiative leptonic decay and the pure-leptonic
decay for the muon are at the same order.

\section{Discussions}

One may see from eqn.(\ref{7}) that the decay rate of 
$B_c \to \gamma \ell \bar \nu$ is
proportional to $f_{B_c}^2$, so one may use
it to measure the decay constant $|f_{B_c}|$. We should note
here that the decay rate eqn.(\ref{7}) is model-dependent, because
the `hadronic piece' in the calculations is model dependent i.e.
the non-relativistic constituent
quark model is used.
However, unlike the $B^\pm \to \gamma \ell \nu$ decays, here b and c quarks
are both heavy.
The long distance contributions can be controlled by the factor of
$\Lambda_{QCD}/m_c$ or $\Lambda_{QCD}/m_b$ thus
they will be able to estimate theoretically \cite{13}.
Therefore to determine the decay constance $f_{B_c}$ from the 
measurements with our formula is still promised.

The similar decays, $B^\pm \to \gamma \ell \nu $, 
have been calculated by many
authors \cite{bu}.
The decay width has been shown in constituent quark model to be
\begin{equation}
\Gamma =\frac{ \alpha G_F^2 |V_{ub}|^2  }{648\pi^2 }
f_{B_u}^2 m_{B_u}^5 /m_u^2 ,
\label{bu}
\end{equation}
where $m_u=350$MeV is the constituent quark mass of u-quark.
To compare the importance of the  $B_c \to \gamma \ell \bar \nu $
and that of the
decay $B^\pm \to \gamma \ell \nu $, let us present
the relative fraction of $\gamma \ell \nu $ final states
coming from different sources $B_c$ and $B_u$
in a high energy production process:
\begin{equation}
\frac{N_{B_c}}{N_{B_u}} =\frac{1}{4}\frac{f(b\to B_c)}{ f(b\to B_u)}
\frac{ |V_{cb}|^2  }{ |V_{ub}|^2  }\frac{f_{B_c}^2}{f_{B_u}^2}
\left(\frac{ m_{B_c}}{ m_{B_u}}\right)^3 \left[ x_b
+x_c \right] \frac{m_u^2}{ m_{B_u}^2} ,
\label{ncu}
\end{equation}
where the factor $f(b\to B_c)$, $f(b\to B_u)$ are the inclusive probability
that a b-quark hadronizes into a $B_c$ meson and a $B_u$ meson respectively.
The $f(b\to B_c)$ is estimated to be a small number of the order 0.1\%,
and the probability $f(b\to B_u)$ is known at LEP with good
accuracy \cite{pdg}:
$$f(b\to B_u)=0.378\pm 0.022.$$
Using the central values of all parameters, one obtains
$$N_{B_c}/N_{B_u}=0.8.$$
This means that the  $\gamma \ell \nu $ final states coming from $B_c$ and
 $B_u$
are at the same order. We expect a similar fraction will also be obtained
at LHC. If one would like to use the radiative decays to determine the decay
constant $f_{B_u}$ (or $f_{B_c}$), 
the background from the corresponding decays of the meson
$B_c$ (or $B_u$) should be considered very carefully.
In B factories at KEK and SLAC, there will be no such problem, since
$B_c$ cannot be produced there.
The energy spectra of the lepton and neutrino for $B_c$ decays in Fig.2
have been emphasized to be different from that for the $B_u$ decays \cite{bu}.
This will be a good feature for experiments to distinguish the 
radiative decays of $B_c$ from that of $B_u$ meson. 

In conclusion, we predict the branching ratios for $B_c \to \gamma 
\ell \bar \nu $ in SM at the order of $10^{-5}$.
With this branching ratio, they are hopeful
detectable at Tevatron and LHC.
When enough $B_c$ meson events are collected, the radiative decays
will be able to provide alternate channels for measuring and/or
`cross-checking' the decay constant $f_{B_c}$ independently
with certain accuracy, provided that
the background from $B_u$ decays is treated well. To enhance the
accuracy of the theoretical predictions, or to estimate theoretical
uncertainty of the calculations, various calculations 
on the $B_c$ radiative decay with different 
models are needed \cite{13}.
 
\section*{Acknowledgment}
We would like to thank G. Kramer for reading the manuscript and for useful
comments. One of the authors (C.D.L) would like to thank the CERN theory 
group for hospitality during his visit.

\end{document}